# Instrument for in-situ orientation of superconducting thin-film resonators used for electron-spin resonance experiments

Andrew Mowry,[1] Yiming Chen,[1,2] James Kubasek,[1] and Jonathan R. Friedman[1,a)]
[1)] *Department of Physics and Astronomy, Amherst College, Amherst, MA 01002-5000*
[2)] *Department of Physics, University of Massachusetts, Amherst, MA 01003*

When used in Electron-Spin Resonance (ESR) measurements, superconducting thin-film resonators must be precisely oriented relative to the external magnetic field in order to prevent the trapping of magnetic flux and the associated degradation of resonator performance. We present a compact design solution for this problem that allows *in-situ* control of the orientation of the resonator at cryogenic temperatures. Tests of the apparatus show that when proper alignment is achieved, there is almost no hysteresis in the field dependence of the resonant frequency.

## I. INTRODUCTION

The use of superconducting thin-film resonators, such as the coplanar waveguide resonator, in electron-spin resonance (ESR) experiments has risen in the past decade because they offer a high quality factor (Q) and small modal volume, which allow for higher sensitivity compared to more traditional methods. The external magnetic fields that must be applied during ESR can degrade Q, as any magnetic-field component perpendicular to the superconducting film surface will lead to Abrikosov vortices, which can produce significant dissipation for high-frequency signals.[1–4] Controlling the vortex formation and behavior is a major challenge in any ESR experiment in which superconducting resonators are used.

Significant progress has been made in efforts to control the vortices using notches[5] or antidots[6,7] in the superconducting film; these pin the vortices, reducing kinetic losses. Alternatively, because the amount of trapped flux is proportional to the component of the magnetic field normal to the film, careful efforts can be made to eliminate this field component, such that the field lies only in the plane of the thin film. Several research groups make use of superconducting striplines or resonators for ESR experiments and some report the alignment issue to require careful attention or consideration.[8–19] Alignment of the resonator plane with the direction of the applied field can be achieved using a vector magnet to control the orientation of the field. Absent such equipment, mechanical alignment of the resonator itself is necessary to prevent flux trapping.

In working with niobium coplanar waveguide resonators, we have found the resonator's behavior to be extremely sensitive to the relative orientation of the resonator and external magnetic field: a half-degree tilt at even 100 Oe renders the apparatus unserviceable. To address this issue, we developed an apparatus to allow for precise *in situ* control of the resonator orientation. Here we present the design of our apparatus and the results of tests of its performance.

[a)] jrfriedman@amherst.edu

## II. CONSTRAINTS

Our ESR apparatus is designed for use in the Quantum Design Physical Properties Measurement System (PPMS). This standard cryogenic platform is convenient for temperature and magnetic field control, but it does not have a vector magnet option, mandating mechanical orientation of the resonator. The PPMS control region – the region of homogeneous and controlled magnetic field and temperature – is $\sim$2.5 cm in diameter and $\sim$5.5 centimeters long, located near the bottom of a cylindrical chamber $\sim$90 cm long. Our niobium coplanar-waveguide resonator sits in this control region, where the field produced by the superconducting magnet is parallel to the sample-chamber axis.

The temperature and magnetic field employed during the experiment impose additional constraints, principally on the materials that can be used. The differential thermal contraction of components that occurs when the system is cooled from room temperature to $\lesssim$ 10 K must be accounted for, and non-magnetic materials and non-superconducting materials must be used in the proximity of the resonator, so as to avoid distorting the field near the resonator. To prevent excessive thermal conduction along the transmission cables going from the room-temperature top plate to the control region, we employ semi-rigid coaxial cables with a stainless steel outer conductor and a silver-plated BeCu inner conductor (Microstock, UT-085B-SS). A collateral benefit of these cables is their high bending stiffness, a property we exploit in our design, as described below.

Our apparatus allows us to precisely rotate the resonator through several degrees *in situ* while maintaining reliable microwave connections between the resonator and the coaxial cables. Below we describe how we achieve this, subject to the above constraints.

## III. DESIGN

We considered several design solutions (including wedges, levers and miniature scissor jacks) to provide the necessary control of the resonator orientation. Ultimately, we chose to use the gear mechanism described below because it provides the finest control given the tight space constraints.

Figure 1 shows a CAD drawing of the lower portion of our ESR probe. The assembly consists of a Nb resonator chip encased in a circuit board assembly (CB - Figure 2), the circuit board holder (CBH - Figure 3), and the gear mechanism (Figure 4), along with the coaxial cables. Exploded views are provided in the Appendix. The basic functioning of this mechanism is as follows. CB is held in CBH by two pins that allow it to pivot about the pins' common axis. The coaxial cables are bent in such a way as to provide a torque that tends to rotate the top of CB towards one side of CBH. The gear mechanism controls a counter torque that allows precise orientation of CB and its embedded resonator. A movie showing the operation of the mechanism can be found in the Appendix (Fig. 10). In the remainder of this section, we provide more detail of each of the components and the mechanism's functioning.

We custom-modified standard SMA launch jacks to maintain good electrical connection with CB in spite of the stress the connectors experience during the angular manipulation. Each brass, nonmagnetic SMA jack is securely attached to the circuit board using two brass L brackets; screws attach each bracket to the outer conductor of the connector and to the ground plane of the copper circuit board. A small piece of soft indium is sandwiched between the center pin of each SMA launch and the central conducting strip of CB to ensure good electrical contact. We avoid using solder for the electrical connections since most common solders are superconducting at the instrument's operating temperatures or are too soft to provide a secure mechanical connection. Silver epoxy was found to be an unreliable alternative.

The semi-rigid coaxial cable loop is made of four sections: Two straight lengths extend from the chamber top plate to the upper working region; these are described in the previous section. Two shorter sections of cable within the working space are made from copper and are terminated with brass SMA connectors that are crimped on to the cable to avoid the use of superconducting solder in the vicinity of the resonator. The shape of the loop in the lower cable is such that, when CB is integrated into the apparatus, the circuit board leans towards the outer coaxial cable (the one passing through the notch in CBH). The cables thus provide the necessary restoring torque for control of the mechanism.

CBH is made of ABS plastic by 3D printing. Its outer diameter is such that it fits snugly inside the sample space at room temperature, although thermal contraction results in somewhat more play at low temperature. CB nestles in the holder's central cavity, which is shaped in the form of a cutout wedge to allow CB to pivot about its lower end. The lower SMA connector of CB has two blind holes into which pins (0-80 screws with the last $\sim 1$ mm of threads machined off) affixed to CBH are inserted, providing a single degree of freedom for the rotation of CB.

The gear mechanism functions, as noted above, to pull the top of CB away from the outer coaxial cable. To achieve this, we designed a central gear (see Fig. 4) that sits in a cartridge on the top of CBH. The gear contains an off-center aperture through which the upper SMA connector of CB passes. The edge of the aperture is in contact with the top SMA jack and as the gear rotates, it causes CB to pivot about the pins inserted into the lower SMA jack. The central gear is driven by a smaller drive gear. The gear ratio of 4:1 allows precise manipulation of the CB angle. The gears are encased within a two-piece cartridge. Each piece of the gear mechanism is machined out of phosphor bronze using a CNC milling machine. The surfaces of the gears are lubricated with Teflon spray (Distec TFL-50 Dry Lube).

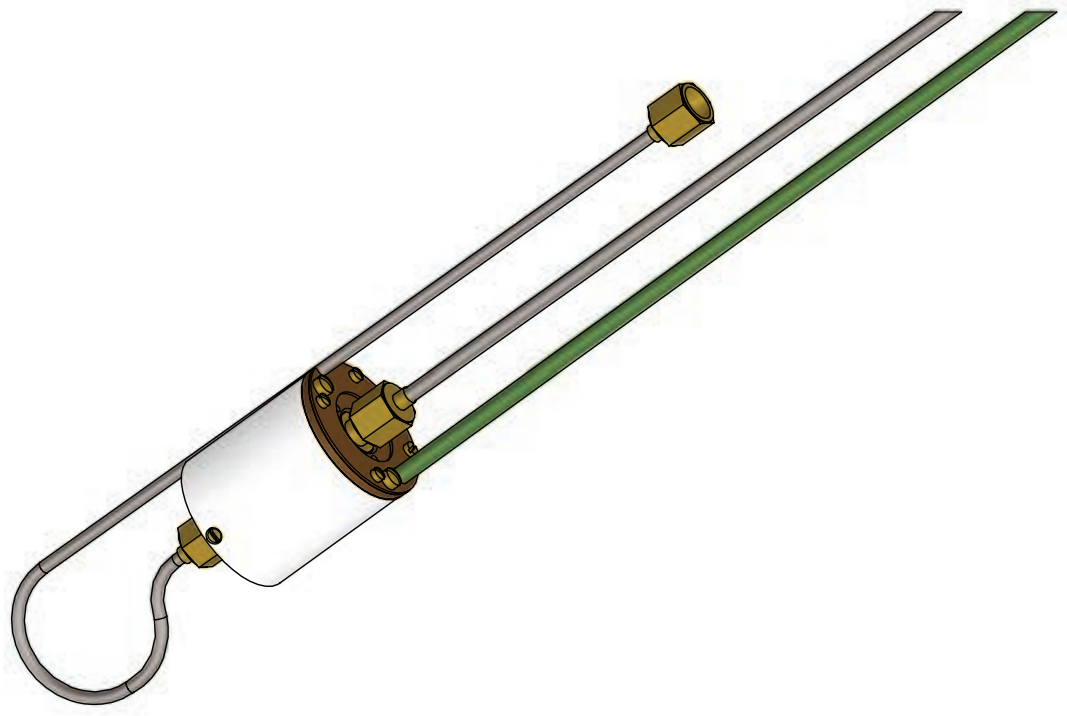

FIG. 1. CAD drawing of the assembled ESR probe. The CBH holds CB. The top SMA connector on CB passes through the gear mechanism, which controls CB's rotation about the bottom pivot pins. The green G10 rod extends from the gear mechanism through the probe top plate, allowing manipulation of the mechanism *in situ*. The semi-rigid coaxial cables also extend through the top plate (not shown). Fig. 6 shows an exploded view.

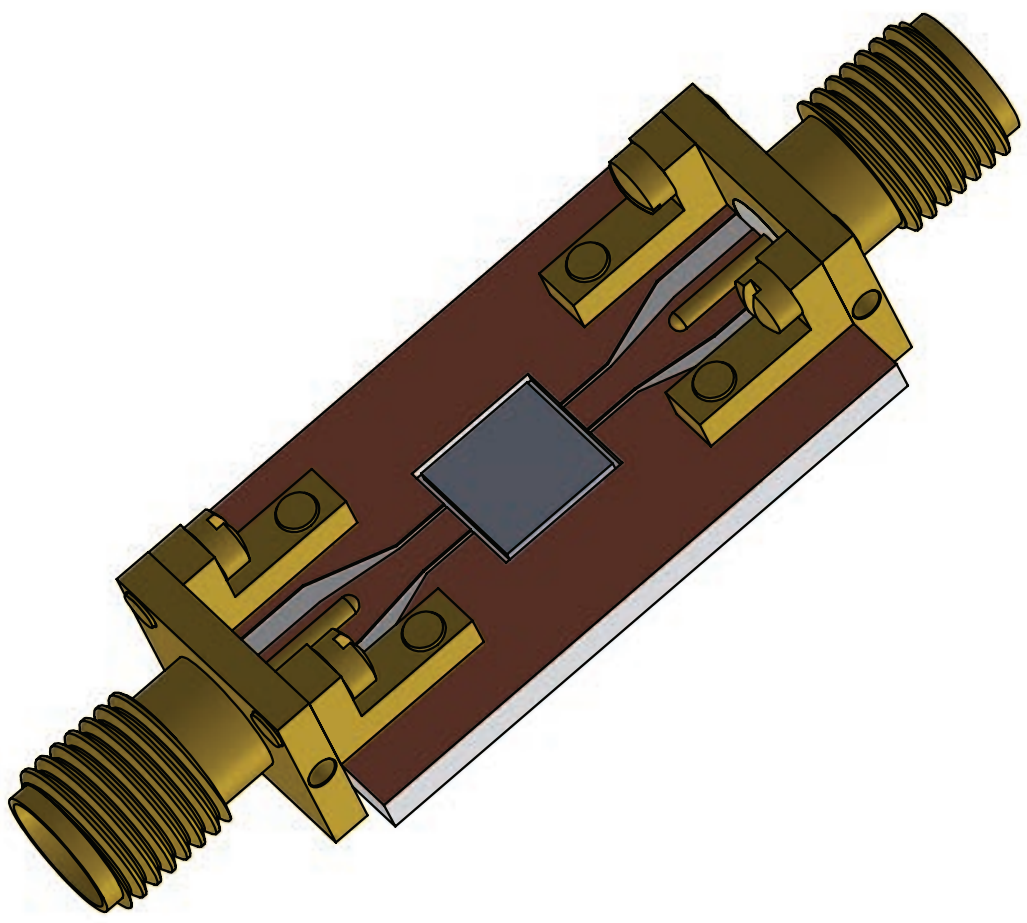

FIG. 2. CAD drawing of the circuit board assembly (CB), including two modified brass SMA connectors affixed with brass brackets. The superconducting coplanar waveguide resonator is located in the center of the circuit board. Fig. 7 shows an exploded view.

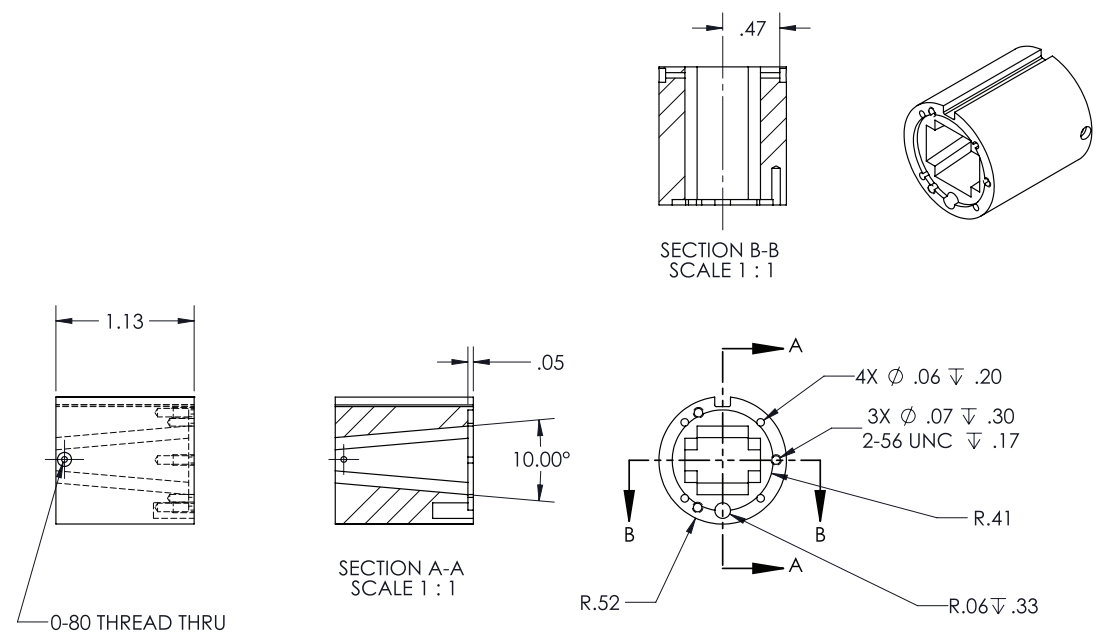

FIG. 3. Technical drawings of the circuit board holder (CBH). The piece was fabricated using a 3D printer. The tapered slot allows CB to pitch through a range of angles.

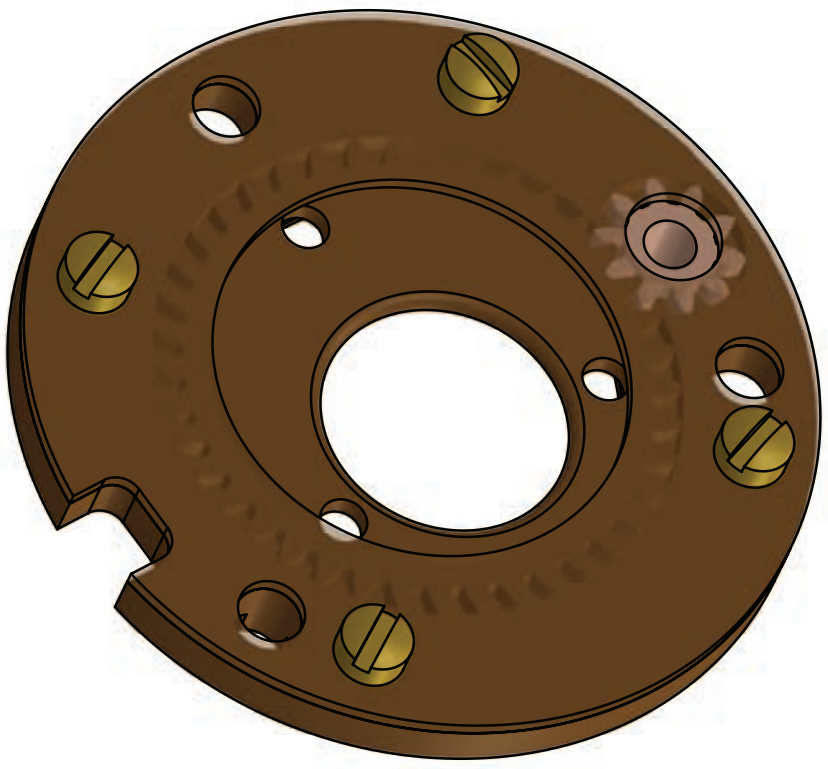

FIG. 4. CAD drawing of the gear mechanism. The top SMA connector of CB passes through the aperature in the larger gear. The gear mechanism is affixed to the top of CBH. The gears can be rotated with respect to the outer casing via a drive shaft threaded into the small gear (hole opposite the notch). Since the rotating central gear forces the aperture to move relative to CBH – and thus relative to the CB's pivot axis – CB's pitch will adjust in a smooth and reproducible manner. The top cover is rendered transparent to make the gears visible. Fig. 8 shows an exploded view.

A drive shaft made from a G10 phenolic rod extends from the gear mechanism up and through the sample chamber top plate, where it terminates in a knob to allow manipulation by the experimenter. The drive shaft typically needs to be disconnected from and reconnected to the gear mechanism between experiments. To this end, the center of the drive gear is threaded, as is the end of the drive shaft. The drive shaft can then easily be screwed into the gear. To allow both clockwise and counterclockwise rotation of the drive gear without unthreading the drive shaft, vacuum grease is applied to the threads: at low temperatures the grease vitrifies, effectively gluing the shaft to the gear. The drive shaft passes through the top plate via a custom hermetic bulkhead connector (Fig. 9) that permits the rod to rotate without compromising vacuum.

## IV. RESULTS

The resonator measured here is a half-wavelength coplanar transmission-line resonator made from a 100-nm-thick Nb film with a designed resonance frequency of $f_{\rm res}$ = 5.2 GHz. It is capacitively coupled to feed lines via 50 $\mu$m-wide gaps at both ends, each of which has a coupling capacitance of 0.24 fF. The chip is mounted into a recess in CB with vacuum grease, and the on-chip feed lines are electrically connected to CB's central strip with Al wire bonds; the ground planes of CB and the resonator chip are similarly wirebonded together. The resonator is characterized using a -12 dBm input signal and a crystal detector to monitor the transmitted power.

The frequency-dependent transmitted power near the fundamental mode of a resonator is measured at 1.8 K and each spectrum is fit to a Lorentzian to extract the resonance frequency $f_{\rm res}$ and quality factor $Q$. Fig. 5(a-c) shows the dependence of $f_{\rm res}$ on the applied field $H$ for three different tilting angles ($\alpha = 5°$, $\lesssim 1°$, and $\sim 0°$, respectively) of the CB plane to the field direction, controlled by the mechanism described above. [Note that each panel has different scales – as $H$ increases, the resonance peak becomes suppressed and fitting of the peak eventually becomes unreliable. The maximum value of $H$ that allows for a reliable fit decreases as $\alpha$ increases.] For $\alpha = 5°$ (Fig. 5(a)), $f_{\rm res}$ drops suddenly when the field is swept up to $\sim$ 50 Oe, which indicates a significant flux-trapping effect. When the field is reduced back from 2 kOe to 0, the data show a strong hysteretic behavior and the original $f_{\rm res}$ is not restored at $H = 0$. The flux trapping issue is essentially resolved after the chip has been carefully oriented parallel to the field (Fig. 5(c)). The resonance frequency follows a smooth curve as the field is ramped up and persists up to a field of 2.2 kOe. Almost no hysteresis is observed when the field is swept back and $f_{\rm res}$ recovers its initial value at $B = 0$, evidence that almost no flux is trapped during the sweep cycle. We found that tilting the chip slightly off the optimal orientation ($\alpha \lesssim 1°$ – Fig. 5(b)) will degrade the resonance curve, indicating that flux trapping is very sensitive to the chip orientation and that accurate control is paramount for the reproducibility of the field dependence of $f_{\rm res}$.

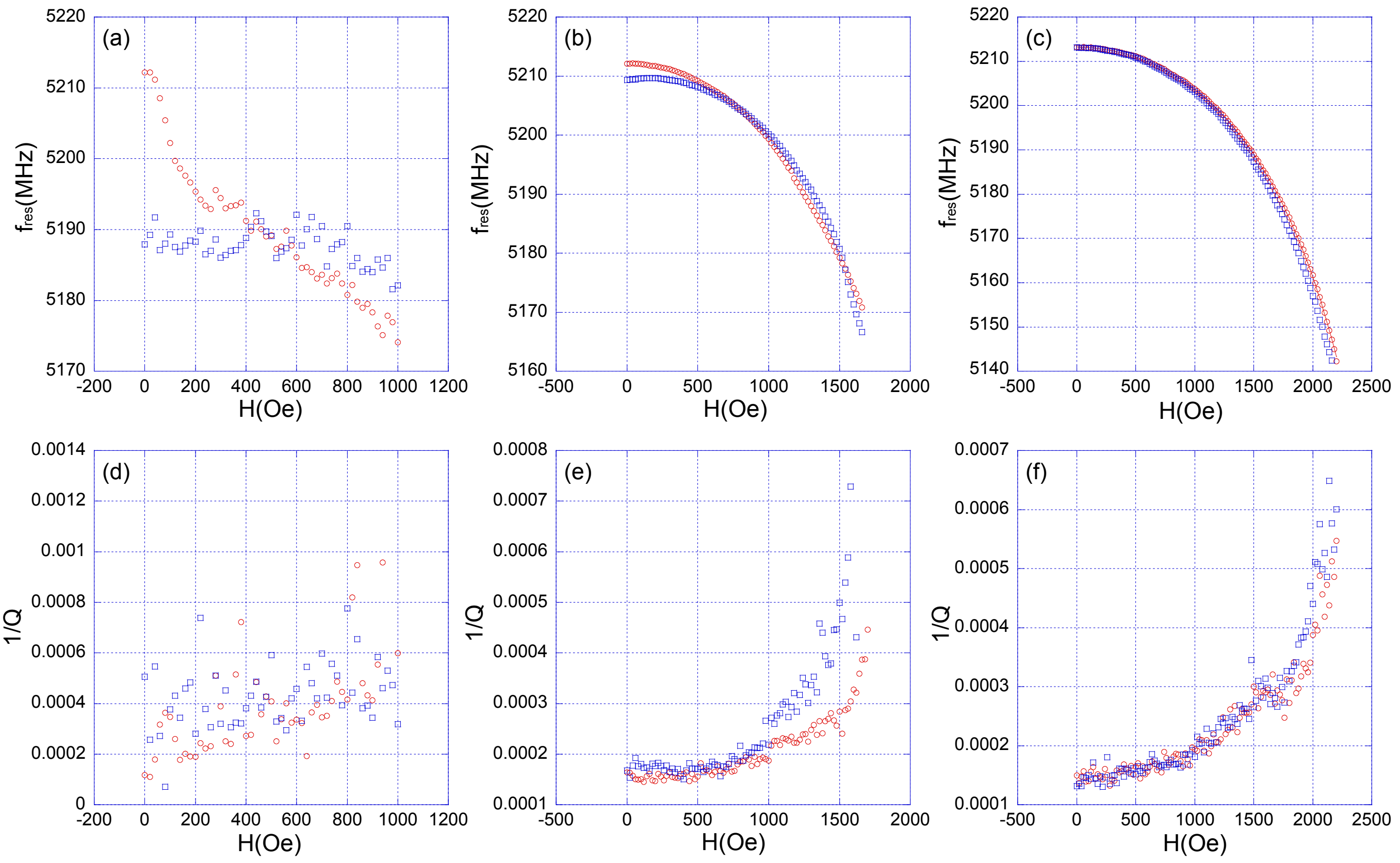

FIG. 5. Resonance frequency as a function of magnetic field as the field is swept up (red circles) and down (blue squares) for three different CB orientations (a) $\alpha \sim 5°$, (b) $\alpha \lesssim 1°$, and (c) $\alpha \sim 0°$ between resonator plane and applied field. The up sweep curve in (c) is fit, as described in the text. (d-f) show $1/Q$ as a function of field for the same orientations as in (a-c), respectively.

Similar conclusions are obtained from the dependence of $1/Q$ on magnetic field. Fig. 5(d-f) show this dependence for the same angles as in Fig. 5(a-c), respectively. For good alignment (Fig. 5(f)), the dependence is smooth with a minimal amount of hysteresis while for larger tilts (Fig. 5(d,e)), the behavior is degraded.

The data in Fig. 5(c) (in the absence of flux trapping) can be understood using standard BCS theory in terms of a field-dependent kinetic inductance of the resonator $L_k$. The resonator can be modeled as an LC circuit with $f_{\rm res} = \frac{1}{\sqrt{LC}}$, where $C$ is the resonator's distributed capacitance and $L$ is the sum of the field-independent geometric inductance $L_m$ and $L_k$. $L_k$ is associated with the magnetic field:[1]

$$L_k(H) = \frac{L_k(0)}{\sqrt{1 - \frac{H^2}{{H_{c||}}^2}}}. \quad (1)$$

We fit the up-sweep data (red) in Fig. 5(c) to $f_{\rm res} = \frac{1}{\sqrt{(L_m + L_k(H))C}}$. The fit is quite good, being difficult to distinguish from the actual data. The fit yields a ratio $\frac{L_k(0)}{L_m} = .04$ and a critical field $H_{c||} = 3465$ Oe. Since $H_{c||} = 2\sqrt{6}\frac{H_c \lambda}{d}$, we determined that the zero-field effective penetration depth to be $\lambda = 35$ nm, using $H_c = 1980$ Oe for niobium and $d = 100$ nm for the thickness of the Nb film in our resonator. This compares well with the bulk value of the London penetration depth, $\lambda_L = 32$ nm, indicating that the superconducting film is of high quality.

## V. DISCUSSION

Our apparatus is designed to work at the temperatures ($\geq 1.8$ K) of the PPMS. Nevertheless, it should be possible to adapt it to work at lower cryogenic temperatures. Many applications of superconducting resonators require use of a dilution refrigerator. We expect many of the materials used to work well at such lower temperatures. Some of the materials, such as indium, will become superconducting at lower temperatures and may need to be replaced by normal metals with similar prop-erties. The mechanical feedthrough for the gear mechanism's drive shaft may need to be carefully designed to prevent air leaks into the fridge. Our design requires a straight connection between the top flange and the sample space, which may not be possible in some dilution refrigerators. Some modification of the design may be necessary to accommodate such constraints. Operation of the mechanism at low temperatures will likely result in appreciable frictional heating. This could be circumvented by performing the alignment while the system is at ~1 K. Once proper alignment has been achieved, there should be little reason to make adjustments upon cooling to lower temperatures.

Aligning the film of the resonator with the magnetic field *in situ* is challenging not only because the small inner diameter of the PPMS chamber limits the working space but also because of the extreme sensitivity of the resonator to an out-of-plane field component. The apparatus that we developed provides a reliable way of tuning the orientation of the resonator so that ESR experiments can be performed in fields up to $\sim$ 2 kOe without trapping flux in the superconducting film. The upper limit of field appears to be a limitation imposed by the thickness of the Nb film used in the fabrication of the resonators and not of the apparatus. Using resonators made with thinner films, we would expect to be able to do reliable experiments at fields above 10 kOe (Ref. 15).

## ACKNOWLEDGMENTS

We thank S. Datta for fabrication of the resonators used in this experiment and S. T. Adams for his work on an early version of the apparatus. We are grateful to P. Chapin for his assistance in the use of the 3D printer. We are indebted to J. Nicholson for his technical help in resonator fabrication, which was carried out, in part, at the Nanotechnology Cleanroom Lab at the University of Massachusetts, Amherst. Support for this work was provided by the National Science Foundation under Grants No. DMR-1006519 and No. DMR-1310135, and by the Amherst College Dean of Faculty.

## APPENDIX: ADDITIONAL DRAWINGS AND MULTIMEDIA ILLUSTRATIONS

Figures 6, 7, and 8 show exploded views of the apparatus, CB, and gear mechanism, respectively, with some details about materials and parts in the associated captions.

Figure 9 shows the custom-made vacuum feedthrough for the drive shaft that permits rotation of the shaft (and control of the mechanism) while maintaining vacuum inside the sample chamber.

Figure 10 shows a short movie illustrating the operation of the mechanism. In addition, the online supplement contains a 3D CAD file (in .step format) of the probe assembly including all components of the apparatus.[20]

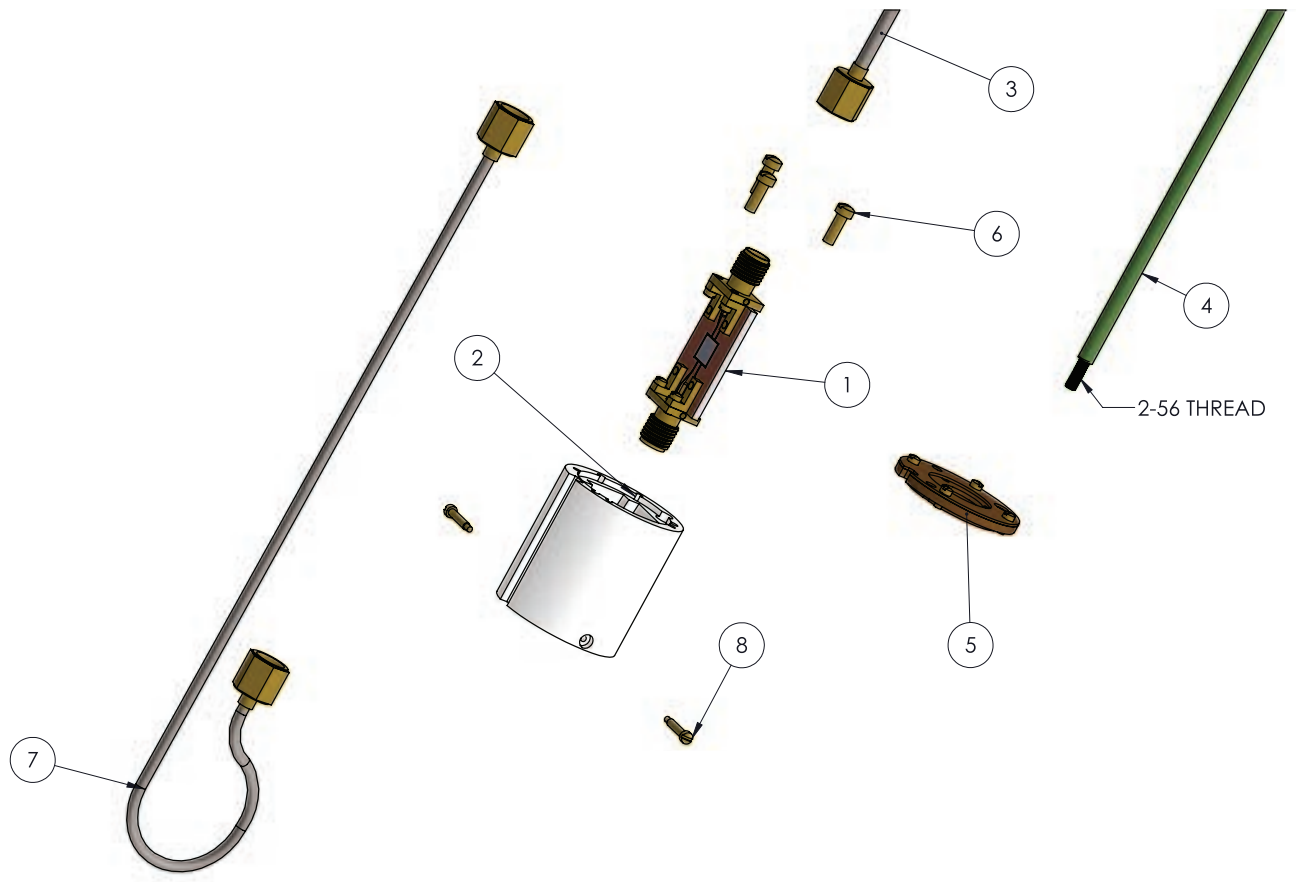

FIG. 6. Exploded view of the apparatus (*c.f.* Fig. 1). ① CB (*c.f.* Fig. 2.); ② CBH (*c.f.* Fig. 3.); ③ & ⑦ non-magnetic, semi-rigid copper coaxial cable (Tek-Stock, UT-085C-LL); ④ threaded G10 drive shaft (.125" diameter, 2-56 thread on last .25"); ⑤ gear mechanism (*c.f.* Figs. 4 and 8); ⑥ 2-56 x .25" brass screw to affix the gear mechanism to CBH; ⑧ pivot pin that affixes CB within CBH.

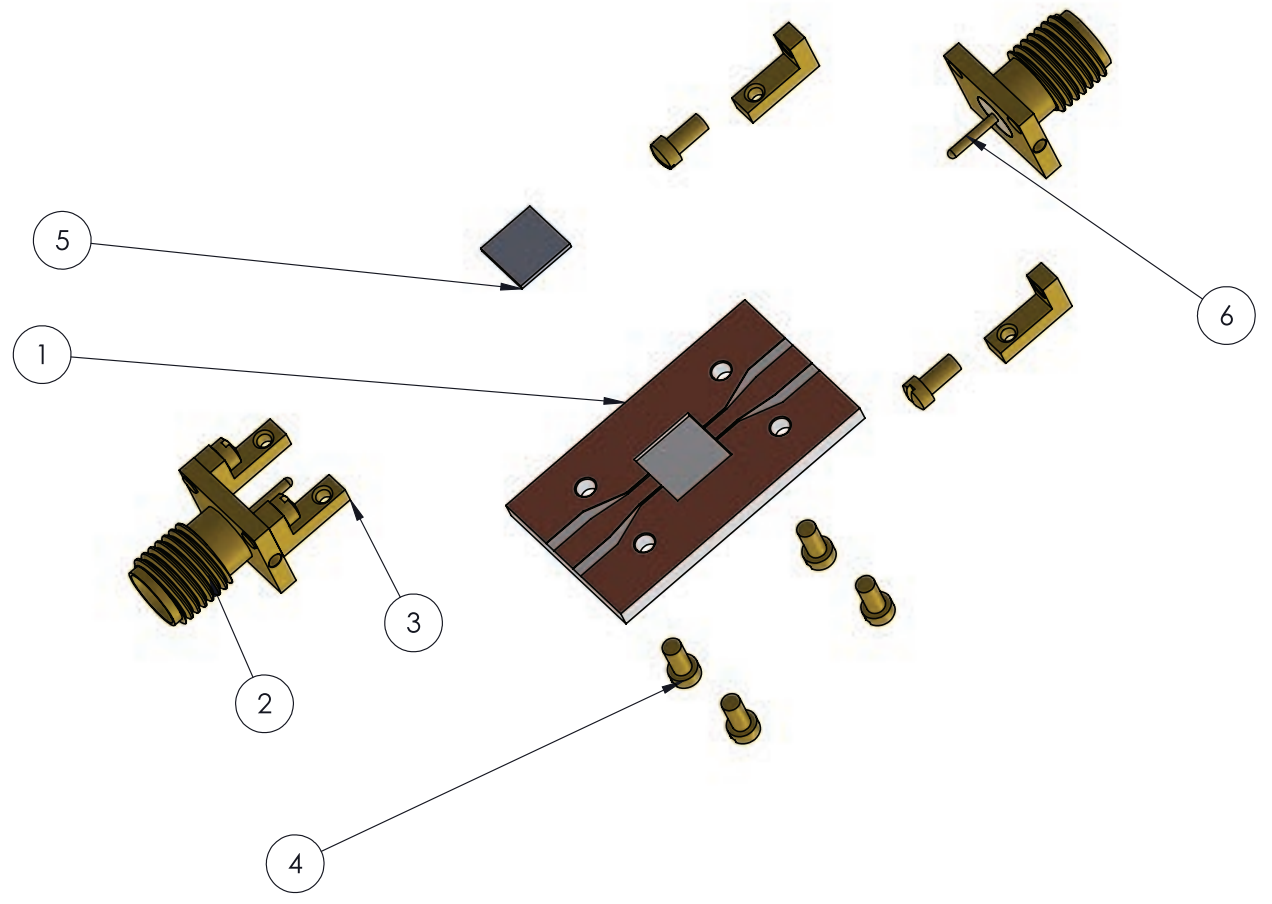

FIG. 7. Exploded view of CB (*c.f.* Fig. 2). ① the circuit board (copper clad ceramic, Rogers Corp., R04350B); ② modified brass SMA connector (Amphenol 132255-12); ③ brass bracket that affixes to the SMA connector and the circuit board; ④ 0-80 brass screw; ⑤ chip containing superconducting (niobium) resonator; ⑥ central conducting pin of the SMA connector, which rests on top of the board's central conducting strip (a small piece of indium (not shown) is sandwiched between the two).

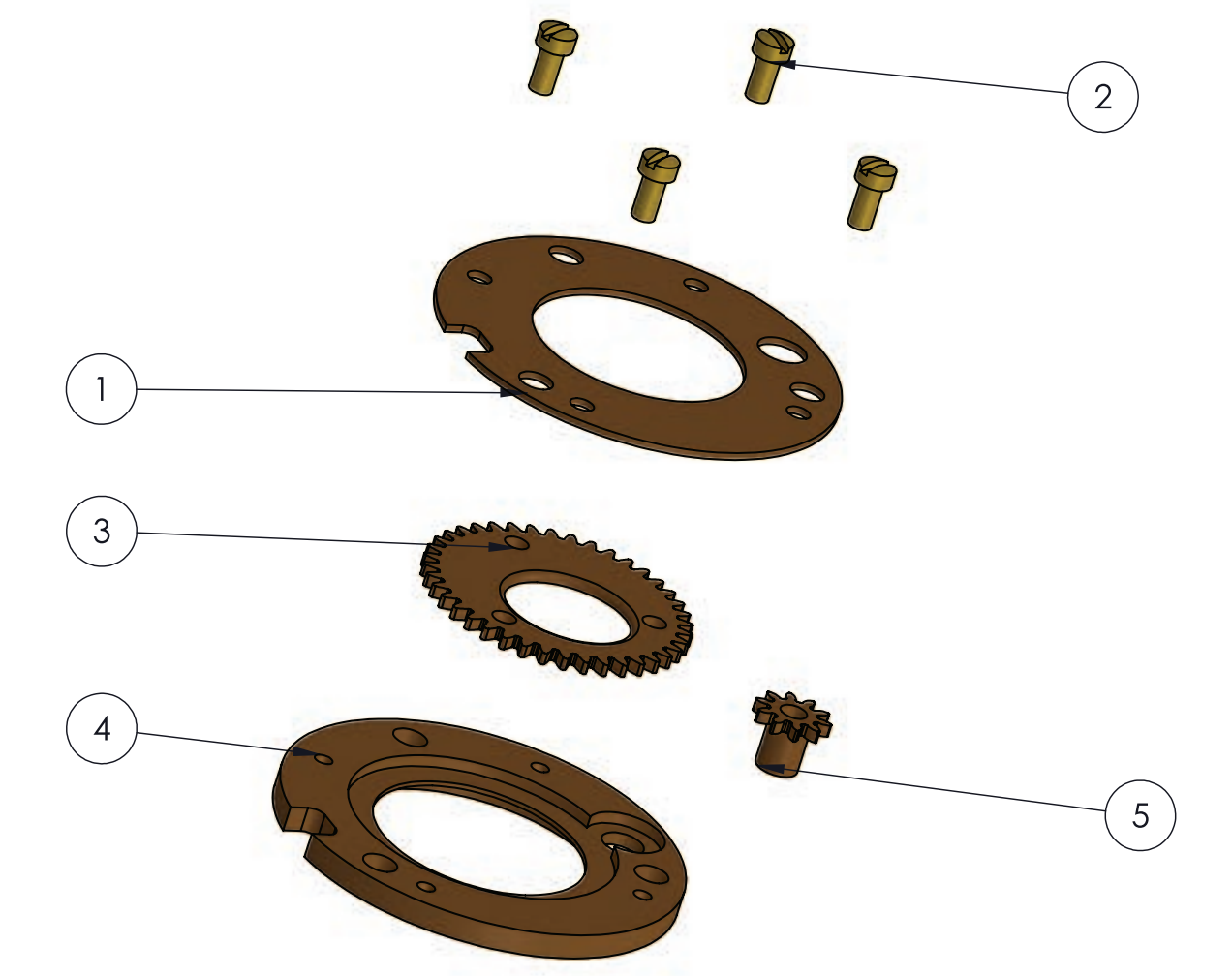

FIG. 8. Exploded view of the gear mechanism (*c.f.* Fig. 4). ① top gear cover plate; ② brass screw (0-80 x .125"); ③ central ring gear with aperture; ④ bottom of gear cartridge; ⑤ drive gear that threads onto the drive shaft (*c.f.* Fig. 6) and interfaces with the central ring gear.

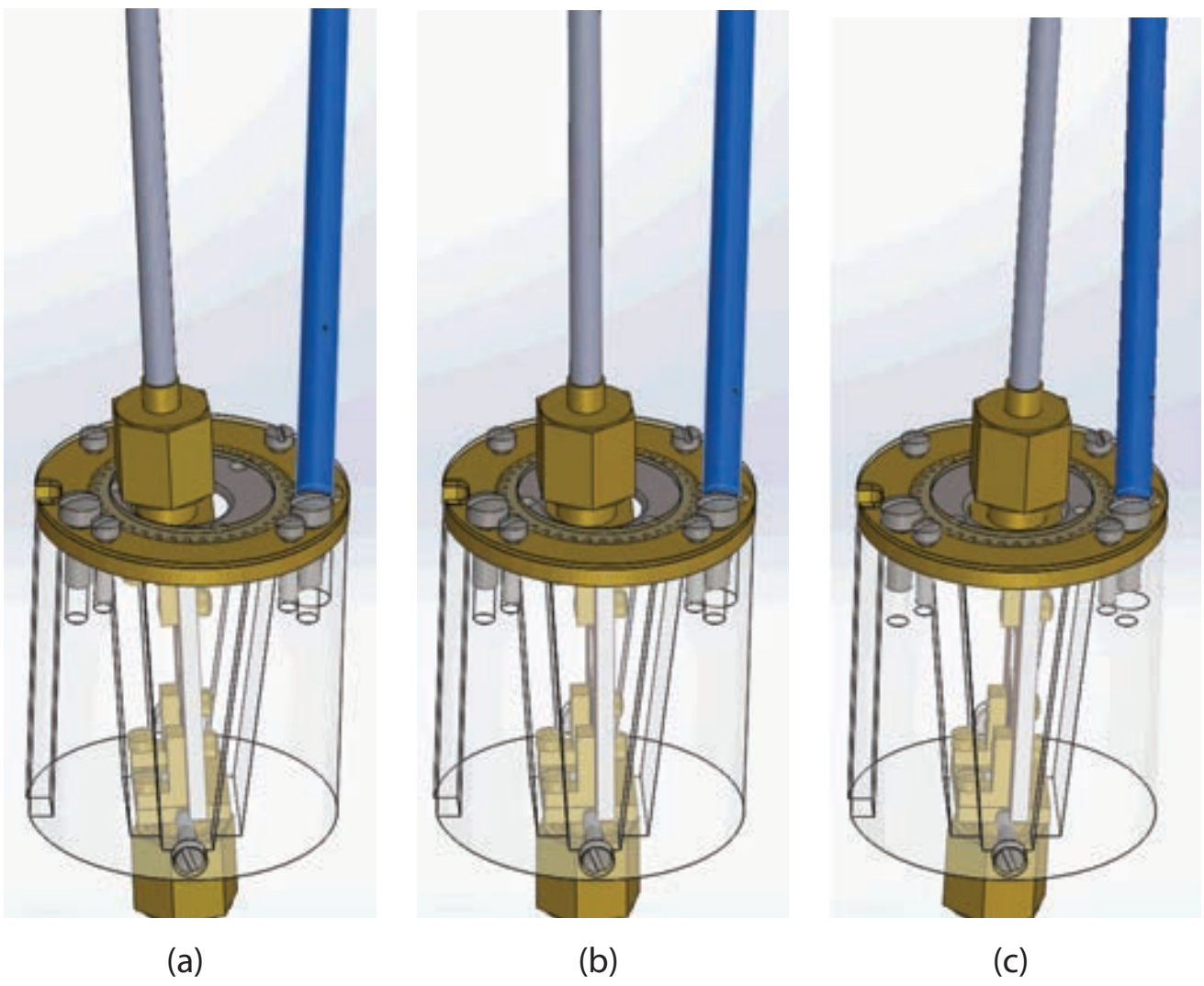

FIG. 10. CAD movie illustrating the operation of the mechanism. The top plate of the gear cartridge and CBH have been rendered transparent for display purposes. (Multimedia view).

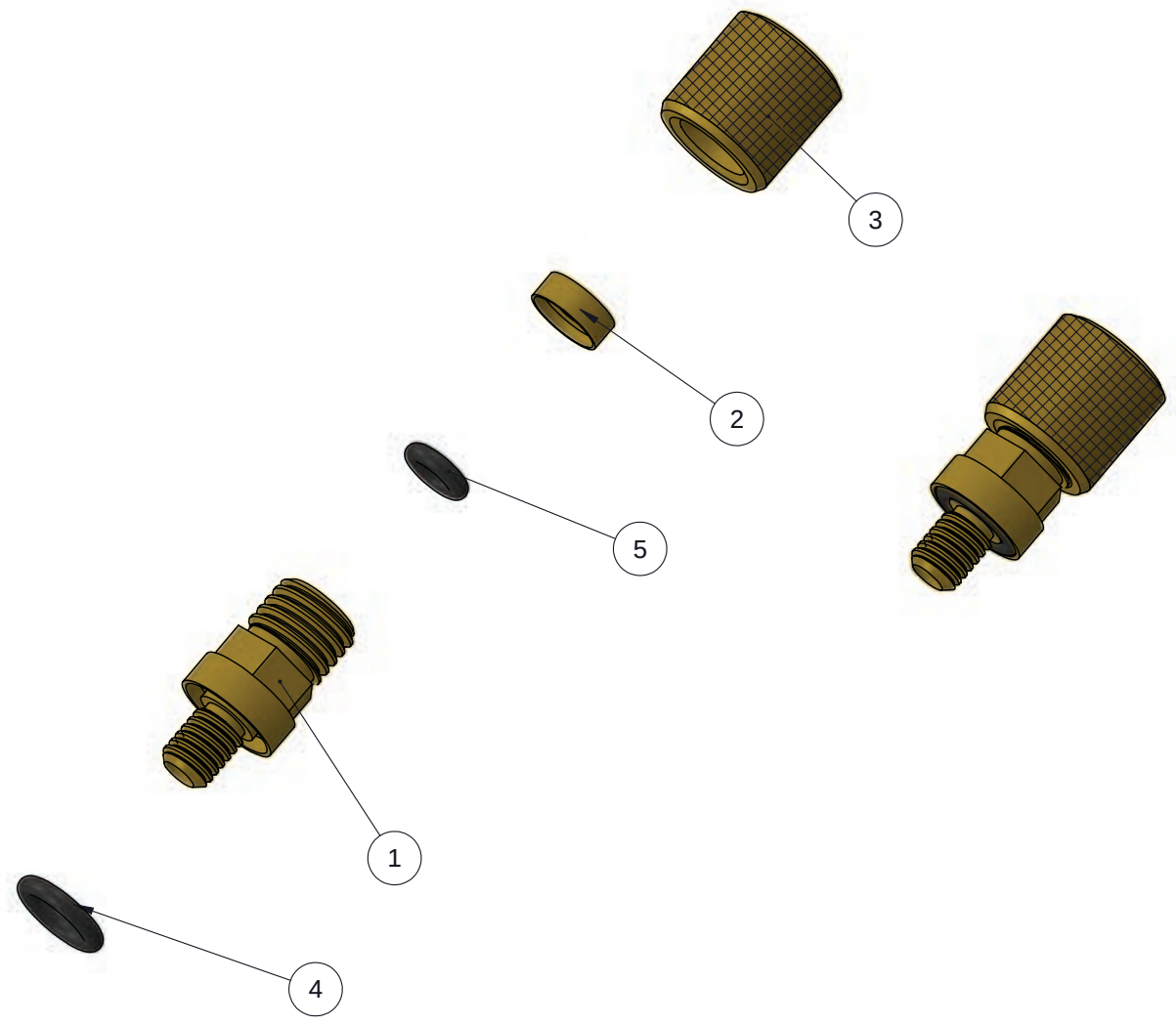

FIG. 9. Exploded and assembled views of the vacuum feedthrough for the drive shaft. ① main connector that screws into the probe top plate; ② centering ring/clamp for O-ring; ③ knurled nut; ④ O-ring for sealing drive shaft (*c.f.* Figs. 1 and 6); ⑤ O-ring for sealing main connector to top plate.